\documentclass[
twocolumn,
]{ceurart}

\usepackage{listings}
\usepackage{cleveref}
\usepackage{hyperref}
\usepackage{enumitem}
\usepackage{amsmath}

\newtheorem{hyp}{Hypothesis}

\newlist{questions}{enumerate}{2}
\setlist[questions,1]{before=\itshape,font=\itshape,label=RQ:,ref=RQ}

\begin{document}

\copyrightyear{2023}
\copyrightclause{Copyright for this paper by its authors. Use permitted under Creative Commons License Attribution 4.0 International (CC BY 4.0).}

\conference{AutomationXP23: Intervening, Teaming, Delegating - Creating Engaging Automation Experiences, CHI ’23, April 23rd, Hamburg, Germany}

\title{On the Perception of Difficulty: Differences between Humans and AI}


\author[1]{Philipp Spitzer}[email=Philipp.Spitzer@kit.edu]
\cormark[1]
\fnmark[1]

\address[1]{Karlsruhe Institute of Technology,
  Kaiserstraße 89-93, Karlsruhe, 76133, Germany}
\address[2]{University of Bayreuth,
Wittelsbacherring 10, Bayreuth, 95444, Germany}
\author[1]{Joshua Holstein}[email=Joshua.Holstein@kit.edu]
\fnmark[1]
\author[1]{Michael V{\"o}ssing}[email=Michael.Voessing@kit.edu]

\author[2]{Niklas K{\"u}hl}[email=Kuehl@uni-bayreuth.de]

\cortext[1]{Corresponding author.}
\fntext[1]{These authors contributed equally.}


\begin{abstract}
With the increased adoption of artificial intelligence (AI) in industry and society,  effective human-AI interaction systems are becoming increasingly important. A central challenge in the interaction of humans with AI is the estimation of difficulty for human and AI agents for single task instances.
These estimations are crucial to evaluate each agent's capabilities and, thus, required to facilitate effective collaboration. So far, research in the field of human-AI interaction estimates the perceived difficulty of humans and AI independently from each other.
However, the effective interaction of human and AI agents depends on metrics that accurately reflect each agent's perceived difficulty in achieving valuable outcomes. Research to date has not yet adequately examined the differences in the perceived difficulty of humans and AI. Thus, this work reviews recent research on the perceived difficulty in human-AI interaction and contributing factors to consistently compare each agent's perceived difficulty, e.g., creating the same prerequisites. Furthermore, we present an experimental design to thoroughly examine the perceived difficulty of both agents and contribute to a better understanding of the design of such systems.

\end{abstract}


\maketitle

\section{Introduction}
In recent decades, technological advances have led to artificial intelligence (AI) applications becoming part of our everyday lives, e.g., when learning a new language \cite{pokrivvcakova2019preparing} or driving autonomous cars \citep{johns2016exploring}. Like many other examples of human-AI interaction, it comes down to appropriately assessing the difficulty of different situations for each agent (human and AI). The consequences for incorrect estimates can range from rejecting such systems, e.g., when the human learner is given too difficult words or grammar without being ready, to potentially severe consequences, e.g., autonomously driving cars on a foggy night. Consequently, it is necessary to estimate each agent's difficulty for an instance adequately.

Further examples of human-AI interaction that draw from an estimation of instance difficulty are \textit{human-AI complementarity} \cite{bansal2021does, hemmer2021human, schemmer2022influence, lubars2019ask, wing2021trustworthy, hemmer2022effect, fugener2022cognitive, steyvers2022bayesian, lai2022human}, \textit{curriculum learning} \citep{bengio2009curriculum, wei2021learn, lotfian2019curriculum}, and \textit{machine teaching} \cite{bengio2009curriculum, wei2021learn, zhu2015machine, zhang2020interactive, singla2014near, spitzer2022training}.
Accurately assessing the difficulty of single instances for both human and AI agents is central to developing these forms of human-AI interaction to fully exploit their complementary capabilities while creating pleasant automation experiences.

By reviewing related literature, we observe different methods and terms, most prominently \textit{uncertainty}, \textit{confidence}, \textit{performance} (e.g., in \cite{taudien2022effect}), for measuring the difficulty of human and AI agents, which is why we aim to delimit our research in the following and create a shared understanding of the relevant terms. Before diving into the frequently used methods, we elaborate on the commonly used terms to describe the difficulty. \textit{Performance} represents the aggregated accuracy over multiple instances for a task or over multiple agents for an instance \cite{taudien2022effect, steyvers2022bayesian}. 

Further, \textit{confidence} and \textit{uncertainty} are often used interchangeably \cite{pouget2016confidence} and serve as a proxy for the difficulty of an instance. However, \citet{pouget2016confidence} argues that these notions are not synonyms. Instead, uncertainty describes the distribution of probabilities over all possible outcomes, while confidence represents the probability that a particular decision is correct. 
When it comes to \textit{difficulty}, one must differentiate between \textit{objective difficulty} and \textit{perceived difficulty}. The former, for instance, can be measured by comparing the number of features of a given task \cite{nicholls1983differentiation}. For the \textit{perceived difficulty}, one must distinguish between task and instance difficulty. On the task level, a common method for human and AI agents depicts the usage of the average performance over multiple instances (for example, in \citet{geirhos2021partial, hemmer2022effect}) to determine the perceived difficulty.

However, on an instance level, the perceived difficulty of human and AI agents is assessed differently.
First, a potential issue arises from an existing gap in access to relevant information. Usually, the AI agent is trained and tested on data drawn from the same distribution, thus having information on the label distribution. However, this is often not the case for humans (e.g., in \cite{geirhos2021partial, peterson2019human, steyvers2022bayesian}). Therefore, it remains unclear whether and how, amongst others, this affects humans' perception of difficulty.
Second, the difficulty of single instances is assessed differently. 
For AI agents, the distribution of the softmax outputs is often used to determine its uncertainty \cite{geirhos2021partial, steyvers2022bayesian}. Contrarily, the human's perceived instance difficulty is often measured by observing the distribution of predictions over groups of humans for single instances or by their average performance for an instance \cite{taudien2022effect, peterson2019human}.
Consequently, individual skills and capabilities of humans are neglected, potentially resulting in poor experiences in human-AI interaction settings \cite{szymanski2021visual}.
As related literature shows, humans have distinct cognitive styles which can affect their perceived difficulty \cite{broverman1960dimensions}. Hence, neglecting their individual traits and generalizing their predictions to determine the perceived difficulty can result in poor estimation for individuals.

As we observe inconsistencies in the measurement of the perceived difficulty between human and AI agents, we outline existing metrics to measure their perceived difficulty as a first step. 
Moreover, we scrutinize methods to compare both agents adequately. Based upon this, we are interested in adequately examining the difference in the perceived difficulty between humans and AI. Therefore, we state the following research question:

\begin{questions}[leftmargin=0.65cm, labelindent=0pt, labelwidth=0em]
    \item What are the differences in the perceived difficulty of humans and AI for single instances?
    \label{researchq}
\end{questions}

To answer this research question, we conduct a literature review to evaluate existing research fields relying on an accurate measurement of the perceived instance difficulty.
Furthermore, we present an experimental design that avoids the previously mentioned inconsistencies. Through our experiment, we want to analyze the perceived difficulty of human and AI agents for single instances, using established metrics like confidence \cite{steyvers2022bayesian} and PVI \cite{ethayarajh2022understanding} adequately. We support our endeavor to establish adequate methods to consistently measure the perceived instance difficulty of human and AI agents with first empirical results based on an existing, public dataset. Overall, with our experiment, we aim to contribute to a better and more integrated understanding of how to adequately compare human and AI agents' perceived difficulty leading to a thorough understanding of the design of human-AI interaction systems.

\section{Related Work}
\label{sec:related}

\subsection{Human-AI Interaction and Instance Difficulty }
With the latest ascent in research on human-AI interaction, the deployment of AI in automated systems is advancing \citep{barboni2010beyond, roto2019engaging}.
Hereby, various forms of human-AI interaction rely on estimating an instance's difficulty for effective collaboration. Following, we outline the three forms of human-AI interaction most relevant to our research: human-AI complementarity, curriculum learning, and machine teaching.

In the field of \textit{human-AI complementarity}, recent research studies complementary team performance---exceeding the performance each agent (human or AI) can achieve on their own \citep{bansal2021does, schemmer2022influence}. In this collaboration, it is crucial to properly delegate tasks to each agent to exploit their complementary capabilities \cite{fugener2022cognitive}. \citet{steyvers2022bayesian} establish a framework to facilitate both human and AI agents' confidence scores to investigate factors that influence complementary capabilities of human-AI collaborations.
\citet{lai2022human} suggest using uncertainty as a measure to delegate tasks between human and AI agents. In the work of \citet{fugener2019collaboration}, the authors evaluate different delegation strategies based on the performance of both agents for single instances. They find that humans' perception of task difficulty differs from the actual task difficulty. \citet{lubars2019ask} investigate, amongst others, the effect of the difficulty of single instances to delegate tasks.

\textit{Curriculum learning} denotes another form of human-AI interaction in which the perceived difficulty is relevant to the overall process. This form of learning is based on human learning and incorporates the idea that the order is crucial in which training instances are presented to a learner \citep{bengio2009curriculum}.
A central aspect of curriculum learning is the assertion of difficulty levels of single instances. \citet{wei2021learn} use the annotator agreement in an image classification task to determine the difficulty of instances.

In the field of \textit{machine teaching}, a human or an AI agent is trained by selecting samples to achieve high learning outcomes \citep{zhu2015machine}. The selection of training instances can be grounded on difficulty estimation. For example, \citet{zhang2020interactive} presents an interactive learning procedure in which crowd workers are trained based on an approximated difficulty for instances. Similarly, \citet{singla2014near} select training instances for learners based on an expected uncertainty measured by an AI agent.

\subsection{Measuring Perceived Difficulty of Humans and AI}

\textit{AI's perceived difficulty.}
In \citet{staahl2020evaluation}, the authors evaluate different metrics to compare the uncertainty of deep learning models. One of these metrics is a Bayesian network-based approach using dropout \citep{gal2016dropout}.
Further, \citet{xu2020theory} present a metric that builds on Shannon entropy \citep{shannon2001mathematical} to compare the difficulty of different datasets. Moreover, \citet{ethayarajh2022understanding} extend this metric, called $\mathcal{V}$-usable information, to apply it to single instances. This metric, the pointwise $\mathcal{V}$-information (PVI), is used to compare the difficulty of single instances with respect to a model family $\mathcal{V}$. According to the authors, PVI, in contrast to related metrics, quantifies the difficulty of single instances accounting for how much information can be extracted beyond the label distribution.

\textit{Human's perceived difficulty.}
Most works focus on estimating the perceived difficulty of humans by aggregating over multiple humans. For example, \citet{peterson2019human} asses the disagreement of two decision-makers. In their work, the authors define the difficulty of a single instance by using the disagreement of crowdsourcing annotators.
To measure the individual perceived difficulty of instances, \citet{steyvers2022bayesian} use a different approach. The authors use the ordinal responses of humans to determine their confidence. Similarly, \citet{biyik2019asking} determine human difficulty by asking participants about their perceived task difficulty.

\section{Empirical Validation Using Public Datasets}
\label{sec:PreFin}

\begin{figure*}[htbp!]
    \centering{\includegraphics[width=\textwidth]{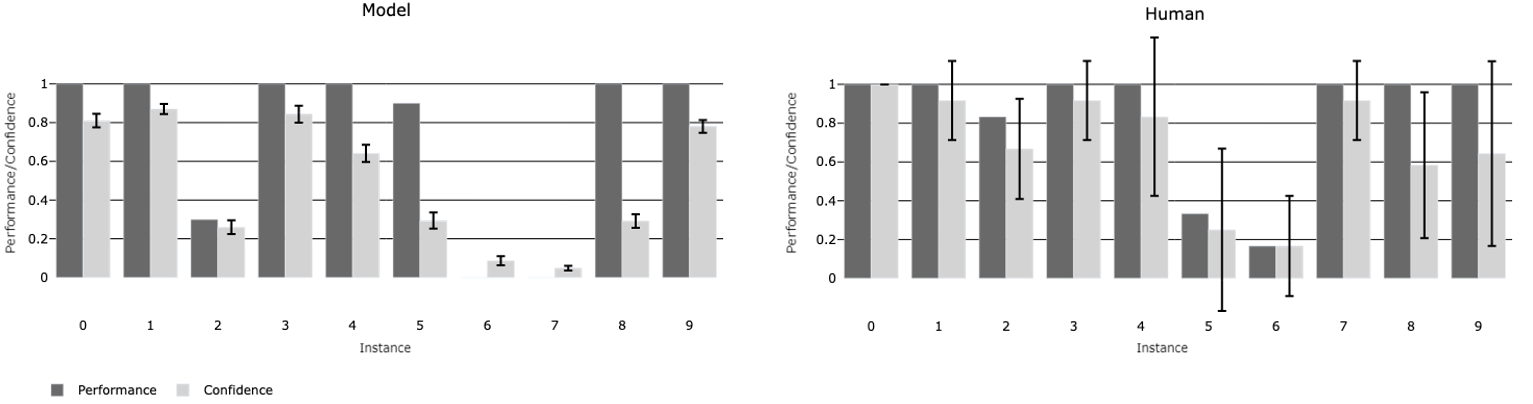}}
	\caption{Comparison of performance and confidence for single instances (top: fine-tuned model, bottom: human).}
	\label{HumanConfidence}       
\end{figure*}

Before our experiment, we examine reports of other studies to investigate the differences in the perceived difficulty of single instances. Therefore, we utilize publicly available datasets, e.g., CIFAR10-H \cite{peterson2019human}, modelvshuman \cite{geirhos2021partial}, or ImageNet-16H \cite{steyvers2022bayesian}.
However, the first two datasets, CIFAR10-H, and modelvshuman do not contain individual human confidence or uncertainty measurements. Instead, the authors of the datasets \cite{geirhos2021partial, peterson2019human} estimate the instance difficulty by aggregating the performance of multiple human annotators for instances. ImageNet-16H is the only dataset containing human difficulty measurements in the form of self-reported confidence levels, e.g., \textit{low}, \textit{medium}, and \textit{high}. To compare these reported confidence levels with the commonly used technique of average instance performance, we transformed the confidence levels to 0 (low), 0.5 (medium), and 1 (high). 

Further, we fine-tune an efficientnet model with the dataset for two epochs and use Monte-Carlo Dropout to receive the perceived confidence of the AI agent. Finally, with the confidence of human and AI agents available, we compare their performance and confidence for single instances.

\Cref{HumanConfidence} illustrates and compares instance performance and confidence for ten randomly sampled instances of the ImageNet-16H dataset. The left part represents the AI agent's output, while the right part shows the human's self-reported confidence. Based on this, we can make several observations.
First, task performance is not necessarily a reliable factor to determine the perceived difficulty of an instance. For example, instances seven to nine have the same performance but differ greatly in their reported confidence. Second, human and AI agents can perceive different instances as easy, e.g., the AI agent has low confidence for instances seven and eight, while the humans have medium to high confidence. Third, the human self-reported confidence scores differ among participants, as can be seen from the standard deviation of confidence. We argue that these observations represent first evidence in the direction of our hypotheses. More specifically, we can see that the average performance of an instance cannot be used to determine the perceived difficulty of an instance for individual humans. Instead, other metrics need to be considered. 

Moreover, the high standard deviation of human confidence for almost all instances indicates that humans differ in their perceived difficulty. Consequently, the diversity of humans must be taken into consideration when designing human-AI interaction systems.

\section{Experimental Design}
\label{sec:PreEx}

Our experiment is based on a mixed-effects model that combines a between-subject and a within-subject design \cite{riefle2022may}.
We follow the notion of existing works and understand confidence as a proxy for difficulty \cite{kompa2021second}. More precisely, we measure the difficulty of the human and the AI agent by two metrics: the commonly used confidence \cite{steyvers2022bayesian} and the PVI score \cite{ethayarajh2022understanding} as a novel metric that considers the label distribution. We measure the confidence of AI agents by Monte-Carlo Dropout \cite{gal2016dropout} and for humans via probabilities, e.g., using a scale between 0\% and 100\%. We use a binary classification task to avoid participants having to assign multiple probabilities. The binary classification allows us to observe one probability, e.g., an image showing a cat with a probability of 80\%, and calculate the complementary probability, e.g., the complementary probability that the image does not represent a cat is 20\%.

\begin{figure}[htbp!]
    \centering{\includegraphics[width=0.47\textwidth]{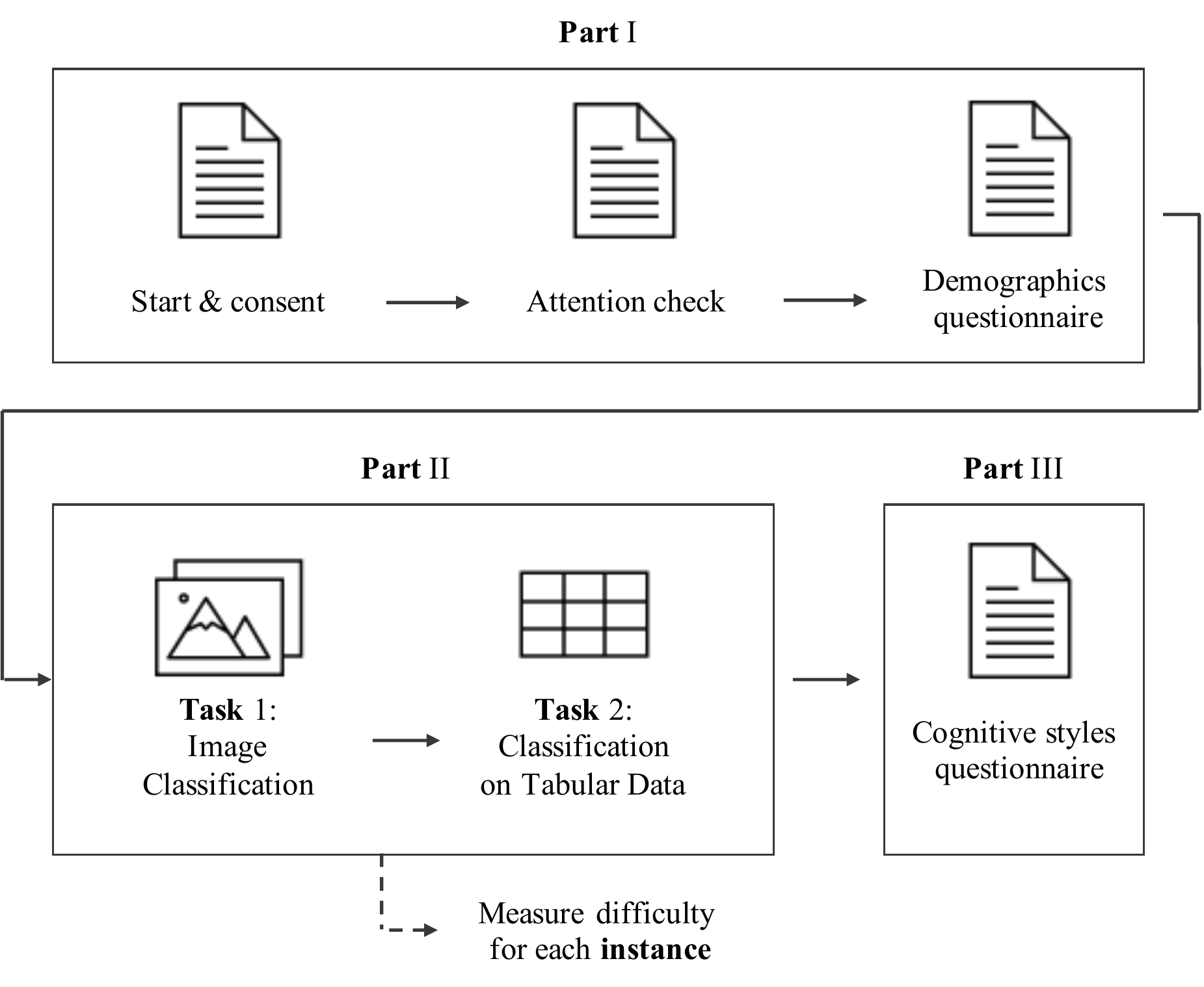}}
	\caption{Design of the study.}
	\label{design}       
\end{figure} 

The preliminary experiment design is illustrated in \Cref{design}. The experiment is composed of three parts. \textit{Part I} includes consent, instructions, and a demographics questionnaire. Next, \textit{Part II} comprises two binary classification tasks---one visual and one textual---and, finally, \textit{Part III} is a questionnaire on cognitive styles. In both tasks, we measure the perceived difficulty of participants and AI for single instances.

In our experiment, we have two treatments. First, as we want a consistent comparison of the perceived difficulty between humans and AI, we must ensure they have access to the relevant information. However, in contrast to humans, the AI agent has access to the label distribution through its training prior to the task. As we want to examine this effect, we show humans the label distribution before conducting the task in one condition. Thus, we hypothesize:
\begin{hyp}
Access to the information on label distribution has an impact on humans' perceived difficulty of single instances.
\label{labeldist}
\end{hyp}

After providing a consistent way to measure the confidence---as a proxy for the perceived difficulty---of the human and the AI agent, we want to examine the differences in their perceived difficulty of instances.
Previous research identified subsets of data on which either human or AI agent has a better performance, e.g., \cite{geirhos2021partial}.
As the performance of human and AI agents is a consequence of the probabilities they assign to each class and, thus, their uncertainty, we argue that the perceived difficulty for an instance can differ even for instances both agents have classified the same. Thus, we hypothesize:

\begin{hyp}
There are instances for which human and AI agents make the same prediction but differ in their perceived difficulty.
\label{hyp1}
\end{hyp}

Within our experiment, we leverage two datasets for the tasks of Part II to compare the perceived difficulty of human and AI agents. Both conditions comprise the same tasks. We chose two different tasks: one visual classification task and one based on tabular data. Research shows the impact of different cognitive styles on participants' task performance (i.e., \citep{broverman1960dimensions, kirby1988verbal, riefle2022influence}). By choosing a visual and a text-based task, we account for participants' different cognitive styles and individual perceptions of difficulty. Accordingly, participants will be asked to conduct a questionnaire in which we determine their cognitive styles.
We assess these styles by using the validated items of \citet{kirby1988verbal} (initially presented by \citet{richardson1977verbalizer}). The items of the cognitive style questionnaire are randomly arranged as suggested by \citet{kirby1988verbal}. All items are measured on a five-point Likert scale.
We hypothesize:
\begin{hyp}
Humans with distinct cognitive styles perceive the difficulty of single instances differently. 
\label{hyp3}
\end{hyp}

\section{Discussion}
\label{sec:Discussion}
In this work, we propose an experimental design to investigate the difference in perceived difficulty between human and AI agents for single instances. To build a foundation, we assess related work and common metrics to estimate instance difficulty. Yet, these studies insufficiently scrutinize consistent difficulty estimations between humans and AI. By first examining a related dataset, we show the discrepancies in difficulty estimation by applying conventional approaches. Thus, we propose an experiment design that paves the way for a broad main study in which we:
(I) Develop a consistent way to measure the perceived difficulty of instances,
(II) Examine the differences in the perceived difficulty of human and AI agents,
(III) Investigate a potential cause in varying perceived difficulty of humans.

Through our main study, we expect to contribute to the ongoing discussion on developing automated and reliable AI agents interacting with humans with diverse skills and capabilities. Moreover, our results will provide guidance not only in research but also in practice on designing human-AI interaction systems. A promising field of research lies ahead.


\bibliography{sample-ceur}

\begin{thebibliography}{39}
\expandafter\ifx\csname natexlab\endcsname\relax\def\natexlab#1{#1}\fi
\providecommand{\url}[1]{\texttt{#1}}
\providecommand{\href}[2]{#2}
\providecommand{\path}[1]{#1}
\providecommand{\DOIprefix}{doi:}
\providecommand{\ArXivprefix}{arXiv:}
\providecommand{\URLprefix}{URL: }
\providecommand{\Pubmedprefix}{pmid:}
\providecommand{\doi}[1]{\href{http://dx.doi.org/#1}{\path{#1}}}
\providecommand{\Pubmed}[1]{\href{pmid:#1}{\path{#1}}}
\providecommand{\bibinfo}[2]{#2}
\ifx\xfnm\relax \def\xfnm[#1]{\unskip,\space#1}\fi
\bibitem[{Pokriv{\v{c}}{\'a}kov{\'a}(2019)}]{pokrivvcakova2019preparing}
\bibinfo{author}{S.~Pokriv{\v{c}}{\'a}kov{\'a}},
\newblock \bibinfo{title}{Preparing teachers for the application of ai-powered
  technologies in foreign language education},
\newblock \bibinfo{journal}{Journal of Language and Cultural Education}
  (\bibinfo{year}{2019}).
\bibitem[{Johns et~al.(2016)Johns, Mok, Sirkin, Gowda, Smith, Talamonti, and
  Ju}]{johns2016exploring}
\bibinfo{author}{M.~Johns}, \bibinfo{author}{B.~Mok},
  \bibinfo{author}{D.~Sirkin}, \bibinfo{author}{N.~Gowda},
  \bibinfo{author}{C.~Smith}, \bibinfo{author}{W.~Talamonti},
  \bibinfo{author}{W.~Ju},
\newblock \bibinfo{title}{Exploring shared control in automated driving},
\newblock in: \bibinfo{booktitle}{2016 11th ACM/IEEE International Conference
  on Human-Robot Interaction (HRI)}, \bibinfo{organization}{IEEE},
  \bibinfo{year}{2016}, pp. \bibinfo{pages}{91--98}.
\bibitem[{Bansal et~al.(2021)Bansal, Wu, Zhou, Fok, Nushi, Kamar, Ribeiro, and
  Weld}]{bansal2021does}
\bibinfo{author}{G.~Bansal}, \bibinfo{author}{T.~Wu},
  \bibinfo{author}{J.~Zhou}, \bibinfo{author}{R.~Fok},
  \bibinfo{author}{B.~Nushi}, \bibinfo{author}{E.~Kamar},
  \bibinfo{author}{M.~T. Ribeiro}, \bibinfo{author}{D.~Weld},
\newblock \bibinfo{title}{Does the whole exceed its parts? the effect of ai
  explanations on complementary team performance},
\newblock in: \bibinfo{booktitle}{Proceedings of the 2021 CHI Conference on
  Human Factors in Computing Systems}, \bibinfo{year}{2021}, pp.
  \bibinfo{pages}{1--16}.
\bibitem[{Hemmer et~al.(2021)Hemmer, Schemmer, V{\"o}ssing, and
  K{\"u}hl}]{hemmer2021human}
\bibinfo{author}{P.~Hemmer}, \bibinfo{author}{M.~Schemmer},
  \bibinfo{author}{M.~V{\"o}ssing}, \bibinfo{author}{N.~K{\"u}hl},
\newblock \bibinfo{title}{Human-ai complementarity in hybrid intelligence
  systems: A structured literature review.},
\newblock \bibinfo{journal}{PACIS}  (\bibinfo{year}{2021}) \bibinfo{pages}{78}.
\bibitem[{Schemmer et~al.(2022)Schemmer, K{\"u}hl, Benz, and
  Satzger}]{schemmer2022influence}
\bibinfo{author}{M.~Schemmer}, \bibinfo{author}{N.~K{\"u}hl},
  \bibinfo{author}{C.~Benz}, \bibinfo{author}{G.~Satzger},
\newblock \bibinfo{title}{On the influence of explainable ai on automation
  bias},
\newblock \bibinfo{journal}{arXiv preprint arXiv:2204.08859}
  (\bibinfo{year}{2022}).
\bibitem[{Lubars and Tan(2019)}]{lubars2019ask}
\bibinfo{author}{B.~Lubars}, \bibinfo{author}{C.~Tan},
\newblock \bibinfo{title}{Ask not what ai can do, but what ai should do:
  Towards a framework of task delegability},
\newblock \bibinfo{journal}{Advances in Neural Information Processing Systems}
  \bibinfo{volume}{32} (\bibinfo{year}{2019}).
\bibitem[{Wing(2021)}]{wing2021trustworthy}
\bibinfo{author}{J.~M. Wing},
\newblock \bibinfo{title}{Trustworthy ai},
\newblock \bibinfo{journal}{Communications of the ACM} \bibinfo{volume}{64}
  (\bibinfo{year}{2021}) \bibinfo{pages}{64--71}.
\bibitem[{Hemmer et~al.(2022)Hemmer, Schemmer, K{\"u}hl, V{\"o}ssing, and
  Satzger}]{hemmer2022effect}
\bibinfo{author}{P.~Hemmer}, \bibinfo{author}{M.~Schemmer},
  \bibinfo{author}{N.~K{\"u}hl}, \bibinfo{author}{M.~V{\"o}ssing},
  \bibinfo{author}{G.~Satzger},
\newblock \bibinfo{title}{On the effect of information asymmetry in human-ai
  teams},
\newblock \bibinfo{journal}{arXiv e-prints}  (\bibinfo{year}{2022})
  \bibinfo{pages}{arXiv--2205}.
\bibitem[{F{\"u}gener et~al.(2022)F{\"u}gener, Grahl, Gupta, and
  Ketter}]{fugener2022cognitive}
\bibinfo{author}{A.~F{\"u}gener}, \bibinfo{author}{J.~Grahl},
  \bibinfo{author}{A.~Gupta}, \bibinfo{author}{W.~Ketter},
\newblock \bibinfo{title}{Cognitive challenges in human--artificial
  intelligence collaboration: Investigating the path toward productive
  delegation},
\newblock \bibinfo{journal}{Information Systems Research} \bibinfo{volume}{33}
  (\bibinfo{year}{2022}) \bibinfo{pages}{678--696}.
\bibitem[{Steyvers et~al.(2022)Steyvers, Tejeda, Kerrigan, and
  Smyth}]{steyvers2022bayesian}
\bibinfo{author}{M.~Steyvers}, \bibinfo{author}{H.~Tejeda},
  \bibinfo{author}{G.~Kerrigan}, \bibinfo{author}{P.~Smyth},
\newblock \bibinfo{title}{Bayesian modeling of human--ai complementarity},
\newblock \bibinfo{journal}{Proceedings of the National Academy of Sciences}
  \bibinfo{volume}{119} (\bibinfo{year}{2022}) \bibinfo{pages}{e2111547119}.
\bibitem[{Lai et~al.(2022)Lai, Carton, Bhatnagar, Liao, Zhang, and
  Tan}]{lai2022human}
\bibinfo{author}{V.~Lai}, \bibinfo{author}{S.~Carton},
  \bibinfo{author}{R.~Bhatnagar}, \bibinfo{author}{Q.~V. Liao},
  \bibinfo{author}{Y.~Zhang}, \bibinfo{author}{C.~Tan},
\newblock \bibinfo{title}{Human-ai collaboration via conditional delegation: A
  case study of content moderation},
\newblock in: \bibinfo{booktitle}{Proceedings of the 2022 CHI Conference on
  Human Factors in Computing Systems}, \bibinfo{year}{2022}, pp.
  \bibinfo{pages}{1--18}.
\bibitem[{Bengio et~al.(2009)Bengio, Louradour, Collobert, and
  Weston}]{bengio2009curriculum}
\bibinfo{author}{Y.~Bengio}, \bibinfo{author}{J.~Louradour},
  \bibinfo{author}{R.~Collobert}, \bibinfo{author}{J.~Weston},
\newblock \bibinfo{title}{Curriculum learning},
\newblock in: \bibinfo{booktitle}{Proceedings of the 26th annual international
  conference on machine learning}, \bibinfo{year}{2009}, pp.
  \bibinfo{pages}{41--48}.
\bibitem[{Wei et~al.(2021)Wei, Suriawinata, Ren, Liu, Lisovsky, Vaickus, Brown,
  Baker, Nasir-Moin, Tomita et~al.}]{wei2021learn}
\bibinfo{author}{J.~Wei}, \bibinfo{author}{A.~Suriawinata},
  \bibinfo{author}{B.~Ren}, \bibinfo{author}{X.~Liu},
  \bibinfo{author}{M.~Lisovsky}, \bibinfo{author}{L.~Vaickus},
  \bibinfo{author}{C.~Brown}, \bibinfo{author}{M.~Baker},
  \bibinfo{author}{M.~Nasir-Moin}, \bibinfo{author}{N.~Tomita}, et~al.,
\newblock \bibinfo{title}{Learn like a pathologist: curriculum learning by
  annotator agreement for histopathology image classification},
\newblock in: \bibinfo{booktitle}{Proceedings of the IEEE/CVF Winter Conference
  on Applications of Computer Vision}, \bibinfo{year}{2021}, pp.
  \bibinfo{pages}{2473--2483}.
\bibitem[{Lotfian and Busso(2019)}]{lotfian2019curriculum}
\bibinfo{author}{R.~Lotfian}, \bibinfo{author}{C.~Busso},
\newblock \bibinfo{title}{Curriculum learning for speech emotion recognition
  from crowdsourced labels},
\newblock \bibinfo{journal}{IEEE/ACM Transactions on Audio, Speech, and
  Language Processing} \bibinfo{volume}{27} (\bibinfo{year}{2019})
  \bibinfo{pages}{815--826}.
\bibitem[{Zhu(2015)}]{zhu2015machine}
\bibinfo{author}{X.~Zhu},
\newblock \bibinfo{title}{Machine teaching: An inverse problem to machine
  learning and an approach toward optimal education},
\newblock in: \bibinfo{booktitle}{Proceedings of the AAAI Conference on
  Artificial Intelligence}, volume~\bibinfo{volume}{29}, \bibinfo{year}{2015},
  pp. \bibinfo{pages}{4083--4087}.
\bibitem[{Zhang et~al.(2020)Zhang, Wang, Meng, and
  Sheng}]{zhang2020interactive}
\bibinfo{author}{J.~Zhang}, \bibinfo{author}{H.~Wang},
  \bibinfo{author}{S.~Meng}, \bibinfo{author}{V.~S. Sheng},
\newblock \bibinfo{title}{Interactive learning with proactive cognition
  enhancement for crowd workers},
\newblock in: \bibinfo{booktitle}{Proceedings of the AAAI Conference on
  Artificial Intelligence}, volume~\bibinfo{volume}{34}, \bibinfo{year}{2020},
  pp. \bibinfo{pages}{540--547}.
\bibitem[{Singla et~al.(2014)Singla, Bogunovic, Bart{\'o}k, Karbasi, and
  Krause}]{singla2014near}
\bibinfo{author}{A.~Singla}, \bibinfo{author}{I.~Bogunovic},
  \bibinfo{author}{G.~Bart{\'o}k}, \bibinfo{author}{A.~Karbasi},
  \bibinfo{author}{A.~Krause},
\newblock \bibinfo{title}{Near-optimally teaching the crowd to classify},
\newblock in: \bibinfo{booktitle}{International Conference on Machine
  Learning}, \bibinfo{organization}{PMLR}, \bibinfo{year}{2014}, pp.
  \bibinfo{pages}{154--162}.
\bibitem[{Spitzer et~al.(2022)Spitzer, K{\"u}hl, and
  Goutier}]{spitzer2022training}
\bibinfo{author}{P.~Spitzer}, \bibinfo{author}{N.~K{\"u}hl},
  \bibinfo{author}{M.~Goutier},
\newblock \bibinfo{title}{Training novices: The role of human-ai collaboration
  and knowledge transfer},
\newblock \bibinfo{journal}{arXiv preprint arXiv:2207.00497}
  (\bibinfo{year}{2022}).
\bibitem[{Taudien et~al.(2022)Taudien, F{\"u}gener, Gupta, and
  Ketter}]{taudien2022effect}
\bibinfo{author}{A.~Taudien}, \bibinfo{author}{A.~F{\"u}gener},
  \bibinfo{author}{A.~Gupta}, \bibinfo{author}{W.~Ketter},
\newblock \bibinfo{title}{The effect of ai advice on human confidence in
  decision-making},
\newblock in: \bibinfo{booktitle}{Proceedings of the 55th Hawaii International
  Conference on System Sciences}, \bibinfo{year}{2022}.
\bibitem[{Pouget et~al.(2016)Pouget, Drugowitsch, and
  Kepecs}]{pouget2016confidence}
\bibinfo{author}{A.~Pouget}, \bibinfo{author}{J.~Drugowitsch},
  \bibinfo{author}{A.~Kepecs},
\newblock \bibinfo{title}{Confidence and certainty: distinct probabilistic
  quantities for different goals},
\newblock \bibinfo{journal}{Nature neuroscience} \bibinfo{volume}{19}
  (\bibinfo{year}{2016}) \bibinfo{pages}{366--374}.
\bibitem[{Nicholls and Miller(1983)}]{nicholls1983differentiation}
\bibinfo{author}{J.~G. Nicholls}, \bibinfo{author}{A.~T. Miller},
\newblock \bibinfo{title}{The differentiation of the concepts of difficulty and
  ability},
\newblock \bibinfo{journal}{Child development}  (\bibinfo{year}{1983})
  \bibinfo{pages}{951--959}.
\bibitem[{Geirhos et~al.(2021)Geirhos, Narayanappa, Mitzkus, Thieringer,
  Bethge, Wichmann, and Brendel}]{geirhos2021partial}
\bibinfo{author}{R.~Geirhos}, \bibinfo{author}{K.~Narayanappa},
  \bibinfo{author}{B.~Mitzkus}, \bibinfo{author}{T.~Thieringer},
  \bibinfo{author}{M.~Bethge}, \bibinfo{author}{F.~A. Wichmann},
  \bibinfo{author}{W.~Brendel},
\newblock \bibinfo{title}{Partial success in closing the gap between human and
  machine vision},
\newblock in: \bibinfo{booktitle}{{Advances in Neural Information Processing
  Systems 34}}, \bibinfo{year}{2021}.
\bibitem[{Peterson et~al.(2019)Peterson, Battleday, Griffiths, and
  Russakovsky}]{peterson2019human}
\bibinfo{author}{J.~C. Peterson}, \bibinfo{author}{R.~M. Battleday},
  \bibinfo{author}{T.~L. Griffiths}, \bibinfo{author}{O.~Russakovsky},
\newblock \bibinfo{title}{Human uncertainty makes classification more robust},
\newblock in: \bibinfo{booktitle}{Proceedings of the IEEE/CVF International
  Conference on Computer Vision}, \bibinfo{year}{2019}, pp.
  \bibinfo{pages}{9617--9626}.
\bibitem[{Szymanski et~al.(2021)Szymanski, Millecamp, and
  Verbert}]{szymanski2021visual}
\bibinfo{author}{M.~Szymanski}, \bibinfo{author}{M.~Millecamp},
  \bibinfo{author}{K.~Verbert},
\newblock \bibinfo{title}{Visual, textual or hybrid: the effect of user
  expertise on different explanations},
\newblock in: \bibinfo{booktitle}{26th International Conference on Intelligent
  User Interfaces}, \bibinfo{year}{2021}, pp. \bibinfo{pages}{109--119}.
\bibitem[{Broverman(1960)}]{broverman1960dimensions}
\bibinfo{author}{D.~M. Broverman},
\newblock \bibinfo{title}{Dimensions of cognitive style.},
\newblock \bibinfo{journal}{Journal of Personality}  (\bibinfo{year}{1960}).
\bibitem[{Ethayarajh et~al.(2022)Ethayarajh, Choi, and
  Swayamdipta}]{ethayarajh2022understanding}
\bibinfo{author}{K.~Ethayarajh}, \bibinfo{author}{Y.~Choi},
  \bibinfo{author}{S.~Swayamdipta},
\newblock \bibinfo{title}{Understanding dataset difficulty with $mathcalv
  $-usable information},
\newblock in: \bibinfo{booktitle}{International Conference on Machine
  Learning}, \bibinfo{organization}{PMLR}, \bibinfo{year}{2022}, pp.
  \bibinfo{pages}{5988--6008}.
\bibitem[{Barboni et~al.(2010)Barboni, Ladry, Navarre, Palanque, and
  Winckler}]{barboni2010beyond}
\bibinfo{author}{E.~Barboni}, \bibinfo{author}{J.-F. Ladry},
  \bibinfo{author}{D.~Navarre}, \bibinfo{author}{P.~Palanque},
  \bibinfo{author}{M.~Winckler},
\newblock \bibinfo{title}{Beyond modelling: an integrated environment
  supporting co-execution of tasks and systems models},
\newblock in: \bibinfo{booktitle}{Proceedings of the 2nd ACM SIGCHI symposium
  on Engineering interactive computing systems}, \bibinfo{year}{2010}, pp.
  \bibinfo{pages}{165--174}.
\bibitem[{Roto et~al.(2019)Roto, Palanque, and Karvonen}]{roto2019engaging}
\bibinfo{author}{V.~Roto}, \bibinfo{author}{P.~Palanque},
  \bibinfo{author}{H.~Karvonen},
\newblock \bibinfo{title}{Engaging automation at work--a literature review},
\newblock in: \bibinfo{booktitle}{Human Work Interaction Design. Designing
  Engaging Automation: 5th IFIP WG 13.6 Working Conference, HWID 2018, Espoo,
  Finland, August 20-21, 2018, Revised Selected Papers 5},
  \bibinfo{organization}{Springer}, \bibinfo{year}{2019}, pp.
  \bibinfo{pages}{158--172}.
\bibitem[{F{\"u}gener et~al.(2019)F{\"u}gener, Grahl, Gupta, and
  Ketter}]{fugener2019collaboration}
\bibinfo{author}{A.~F{\"u}gener}, \bibinfo{author}{J.~Grahl},
  \bibinfo{author}{A.~Gupta}, \bibinfo{author}{W.~Ketter},
  \bibinfo{title}{Collaboration and delegation between humans and AI: An
  experimental investigation of the future of work},
  \bibinfo{publisher}{Erasmus Research Institute of Management (ERIM)},
  \bibinfo{year}{2019}.
\bibitem[{St{\aa}hl et~al.(2020)St{\aa}hl, Falkman, Karlsson, and
  Mathiason}]{staahl2020evaluation}
\bibinfo{author}{N.~St{\aa}hl}, \bibinfo{author}{G.~Falkman},
  \bibinfo{author}{A.~Karlsson}, \bibinfo{author}{G.~Mathiason},
\newblock \bibinfo{title}{Evaluation of uncertainty quantification in deep
  learning},
\newblock in: \bibinfo{booktitle}{Information Processing and Management of
  Uncertainty in Knowledge-Based Systems: 18th International Conference, IPMU
  2020, Lisbon, Portugal, June 15--19, 2020, Proceedings, Part I 18},
  \bibinfo{organization}{Springer}, \bibinfo{year}{2020}, pp.
  \bibinfo{pages}{556--568}.
\bibitem[{Gal and Ghahramani(2016)}]{gal2016dropout}
\bibinfo{author}{Y.~Gal}, \bibinfo{author}{Z.~Ghahramani},
\newblock \bibinfo{title}{Dropout as a bayesian approximation: Representing
  model uncertainty in deep learning},
\newblock in: \bibinfo{booktitle}{international conference on machine
  learning}, \bibinfo{organization}{PMLR}, \bibinfo{year}{2016}, pp.
  \bibinfo{pages}{1050--1059}.
\bibitem[{Xu et~al.(2020)Xu, Zhao, Song, Stewart, and Ermon}]{xu2020theory}
\bibinfo{author}{Y.~Xu}, \bibinfo{author}{S.~Zhao}, \bibinfo{author}{J.~Song},
  \bibinfo{author}{R.~Stewart}, \bibinfo{author}{S.~Ermon},
\newblock \bibinfo{title}{A theory of usable information under computational
  constraints},
\newblock \bibinfo{journal}{arXiv preprint arXiv:2002.10689}
  (\bibinfo{year}{2020}).
\bibitem[{Shannon(2001)}]{shannon2001mathematical}
\bibinfo{author}{C.~E. Shannon},
\newblock \bibinfo{title}{A mathematical theory of communication},
\newblock \bibinfo{journal}{ACM SIGMOBILE mobile computing and communications
  review} \bibinfo{volume}{5} (\bibinfo{year}{2001}) \bibinfo{pages}{3--55}.
\bibitem[{B{\i}y{\i}k et~al.(2019)B{\i}y{\i}k, Palan, Landolfi, Losey, and
  Sadigh}]{biyik2019asking}
\bibinfo{author}{E.~B{\i}y{\i}k}, \bibinfo{author}{M.~Palan},
  \bibinfo{author}{N.~C. Landolfi}, \bibinfo{author}{D.~P. Losey},
  \bibinfo{author}{D.~Sadigh},
\newblock \bibinfo{title}{Asking easy questions: A user-friendly approach to
  active reward learning},
\newblock \bibinfo{journal}{arXiv preprint arXiv:1910.04365}
  (\bibinfo{year}{2019}).
\bibitem[{Riefle et~al.(2022)Riefle, Benz, and Tomar}]{riefle2022may}
\bibinfo{author}{L.~Riefle}, \bibinfo{author}{C.~Benz},
  \bibinfo{author}{T.~Tomar},
\newblock \bibinfo{title}{“may i help you?”: Exploring the effect of
  individuals’ self-efficacy on the use of conversational agents},
\newblock \bibinfo{journal}{ICIS 2022 Proceedings}  (\bibinfo{year}{2022}).
\bibitem[{Kompa et~al.(2021)Kompa, Snoek, and Beam}]{kompa2021second}
\bibinfo{author}{B.~Kompa}, \bibinfo{author}{J.~Snoek}, \bibinfo{author}{A.~L.
  Beam},
\newblock \bibinfo{title}{Second opinion needed: communicating uncertainty in
  medical machine learning},
\newblock \bibinfo{journal}{NPJ Digital Medicine} \bibinfo{volume}{4}
  (\bibinfo{year}{2021}) \bibinfo{pages}{4}.
\bibitem[{Kirby et~al.(1988)Kirby, Moore, and Schofield}]{kirby1988verbal}
\bibinfo{author}{J.~R. Kirby}, \bibinfo{author}{P.~J. Moore},
  \bibinfo{author}{N.~J. Schofield},
\newblock \bibinfo{title}{Verbal and visual learning styles},
\newblock \bibinfo{journal}{Contemporary educational psychology}
  \bibinfo{volume}{13} (\bibinfo{year}{1988}) \bibinfo{pages}{169--184}.
\bibitem[{Riefle et~al.(2022)Riefle, Hemmer, Benz, V{\"o}ssing, and
  Pries}]{riefle2022influence}
\bibinfo{author}{L.~Riefle}, \bibinfo{author}{P.~Hemmer},
  \bibinfo{author}{C.~Benz}, \bibinfo{author}{M.~V{\"o}ssing},
  \bibinfo{author}{J.~Pries},
\newblock \bibinfo{title}{On the influence of cognitive styles on users'
  understanding of explanations},
\newblock \bibinfo{journal}{ICIS 2022 Proceedings}  (\bibinfo{year}{2022}).
\bibitem[{Richardson(1977)}]{richardson1977verbalizer}
\bibinfo{author}{A.~Richardson},
\newblock \bibinfo{title}{Verbalizer-visualizer: a cognitive style dimension.},
\newblock \bibinfo{journal}{Journal of mental imagery}  (\bibinfo{year}{1977}).

\end{thebibliography}

\appendix

\end{document}